\documentclass{article}

\usepackage{graphicx}
\usepackage{amssymb}
\usepackage{epstopdf}
\usepackage[utf8]{inputenc} % allow utf-8 input
\usepackage[T1]{fontenc}    % use 8-bit T1 fonts
\usepackage{hyperref}       % hyperlinks
\usepackage{url}            % simple URL typesetting
\usepackage{booktabs}       % professional-quality tables
\usepackage{amsfonts}       % blackboard math symbols
\usepackage{nicefrac}       % compact symbols for 1/2, etc.
\usepackage{microtype}      % microtypography
\usepackage{xcolor}         % colors

\newcommand{\bs}{\mathbf{s}}

\title{A random energy approach to deep learning}

\author{%
  Rongrong Xie \\
  \small{Key Laboratory of Quark and Lepton Physics (MOE) and Institute of Particle Physics,} \\
  \small{Central China Normal University (CCNU), Wuhan, China} \\
  and \\
  Matteo~Marsili\thanks{marsili@ictp.it} \\
    \small{Quantitative Life Sciences Section}\\
    \small{The Abdus Salam International Centre for Theoretical Physics,
  34151 Trieste, Italy}}

\begin{document}

\maketitle

\begin{abstract}
We study a generic ensemble of deep belief networks which is parametrized by the distribution of energy levels of the hidden states of each layer. 
We show that, within a random energy approach, statistical dependence can propagate from the visible to deep layers only if each layer is tuned close to the critical point during learning. As a consequence,  efficiently trained learning machines are characterised by a broad distribution of energy levels. The analysis of Deep Belief Networks and Restricted Boltzmann Machines on different datasets confirms these conclusions.
\end{abstract}

\section{Introduction}

The study of ensembles of random systems can provide several insights on the properties of complex systems, such as heavy ions~\cite{wigner1993characteristic}, ecologies~\cite{may1972will}, disordered materials~\cite{mezard1987spin}, satisfiability in computer science~\cite{monasson1999determining} and machine learning~\cite{zdeborova2016statistical}. Indeed the collective behaviour of a system composed of many interacting degrees of freedom often does not depend on the specific realisation of the wiring of the interactions, but only on the statistical properties of the resulting energy landscape. In these circumstances, any realisation of a random system that shares the same statistical properties enjoys the same ``typical" collective behaviour. 

The Random Energy Model (REM)~\cite{derrida1981random} is probably the simplest exemplar of this approach. It makes minimal assumptions on the network of interactions, because interactions of any order can occur among the variables~\cite{derrida1981random}. It features a phase transition between a random (high temperature) phase and a low temperature frozen phase, which reproduces the gross features of more complex systems such as spin glasses. 

Here we adopt the same approach to study systems that form an internal representation of a complex environment, such as deep belief networks (DBN). A DBN is composed of a stack of layers of variables, which interact only with variables in neighbouring layers, as shown in Fig.~\ref{fig_OLMs} a). The bottom layer is in contact with the data that the network seeks to learn, that is akin to the environment for a physical system. Typically the data has  a non-trivial hidden structure of statistical dependencies, which the learning machine aims at extracting.

During training, the interaction strengths are adjusted in such a way that the distribution of internal states of the machine form a representation of the data, in the sense that when a state is drawn from this distribution and it is propagated to the bottom layer it reproduces the distribution of the data with which the machine had been trained.  The distribution of internal states after training depends on the structure of the data, on the machine's architecture and on the initial condition for the weights (which is generally assumed random, see e.g.~\cite{hinton2012practical}), and is generally characterised by a complex energy landscape, with effective interactions between variables of arbitrary order. In this paper, we're going to consider this system as a  draw from an ensemble of random systems in such a way that each layer can be thought of as a REM.  This allows us to characterise the properties of learned representations within a rather general framework. Our main conclusion is that, in order for the information on the data's statistical dependencies to propagate to the deep layers, each layer has to be tuned to the critical point. 

There is extensive evidence that learning efficiency and criticality are related~\cite{langton1990computation,bertschinger2004real,roli2018dynamical}. Critical models that are close to a phase transition or the edge of chaos are much better learners and generalisers than non-critical ones. The brain itself is conjectured to operate close to a critical state~\cite{beggs2008criticality,plenz2021self}. Criticality affords several benefits to information processing~\cite{shew2013functional} but there is no agreed general rationale that explains the ubiquitously observed relation between criticality and efficiency in learning. For deep learning, in particular, Schoenholtz {\em et al.}~\cite{schoenholz2016deep} reach conclusions similar to ours. The study deep (deterministic) neural networks with random Gaussian weights and argue that ``only when the statistics of weights and variances is such that the network is close to the edge of chaos the signal propagates from the bottom to deep layers''. Yet their results apply to untrained networks with random weights. Our more general approach reaches the same conclusion for well trained deep (probabilistic) neural networks, irrespective of the statistics of the weights, of the data and of the architecture. 
A similar REM approach can be extended to generic systems who form internal representations of their fluctuating environment, thereby offering a rationale for the observation of criticality in living systems~\cite{MoraBialek}. 

In what follows we shall first set the stage in Section \ref{sec:frame} by introducing the main problem and the notations. Next we shall focus on a random energy description of learning machines in Section \ref{sec:remfit}. We shall end with a final discussion. 

\section{The general framework}
\label{sec:frame}

\begin{figure}
  \centering
\includegraphics[width=0.8\textwidth,angle=0]{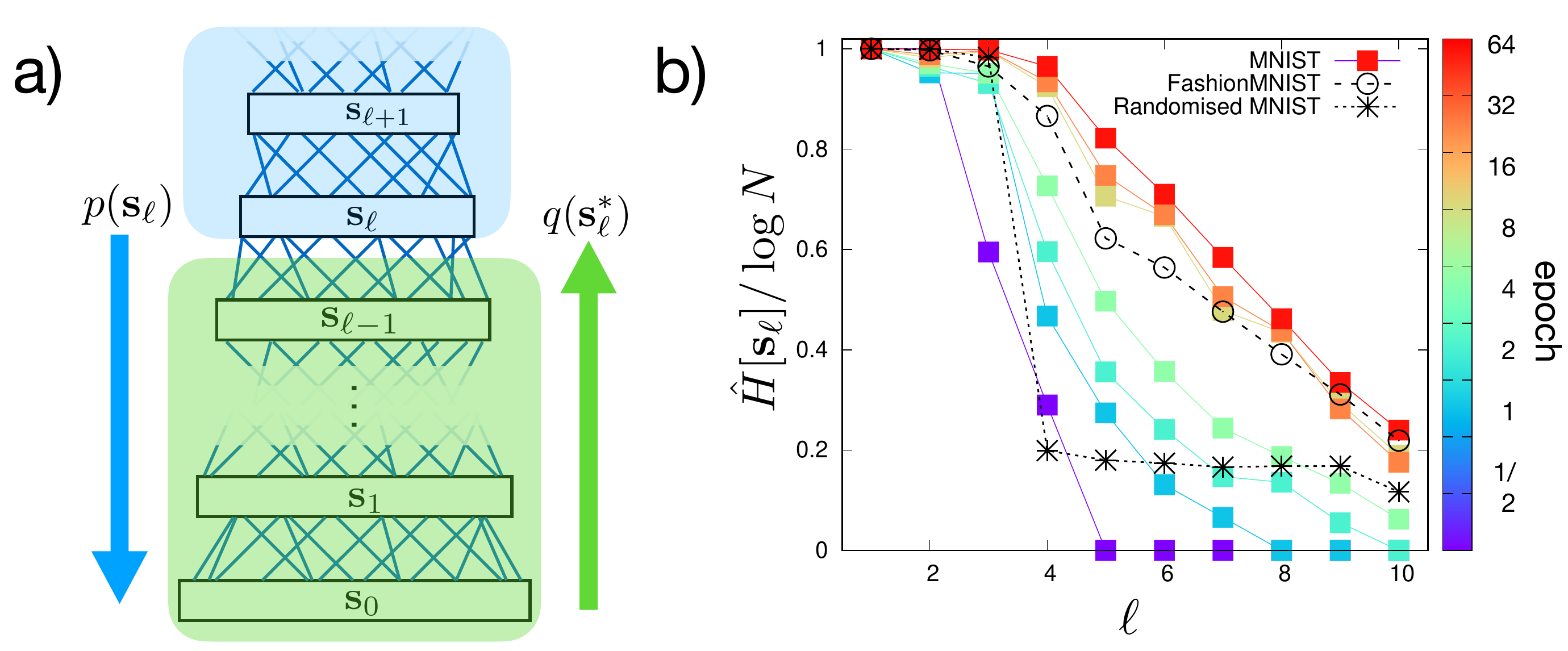}
  \caption{\label{fig_OLMs} {\em a)} typical architecture of a deep belief network. The statistical properties of an internal layer (shaded blue upper part) depend on the shallower layers and with the data (shaded green lower part) which is the analogue of the environment in which the internal layer interacts with. 
  The two arrows symbolically indicate the fact that {\em i)} the most likely states $\bs_\ell^*(\bs_0)$ for each data point $\bs_0$ in the dataset induce a distribution $q(\bs^*_\ell)$ in the hidden states of the DBN, and that {\em ii)} the learned representation $p(\bs_\ell)$ can be used as a generative process. {\em b)} Entropy $\hat H[\bs_\ell]$ of the internal layer $\ell$, as a function of the depth $\ell$ of the layer for a DBN trained on the MNIST dataset with $n_\ell=500, 250, 120, 60, 30, 25, 20, 15, 10$ and $5$, for $\ell=1,2,\ldots, 9$. The entropies of the representations at different stages of training are shown in different colours (see colour bar on the right) for the MNIST dataset. Data for a DBN with the same architecture trained on the Fashion MNIST and on randomised MNIST datasets are shown with open circles ($\circ$) and asterisks ($\ast$), respectively (after 64 epochs of training). The randomised MNIST dataset is obtained randomising the positions of pixels within each image.}
\end{figure}

%In order to make the Random Energy Model~\cite{derrida1981random}, we address here 
In order to set the stage, let us discuss the generic properties of a systems with an architecture similar to that of deep belief networks (DBN)~\cite{roudi2015learning}, such as that shown in Fig.~\ref{fig_OLMs} a). We denote by $\ldots, \bs_{\ell+1},\bs_\ell,\bs_{\ell-1},\ldots,\bs_1$ the internal state of the different layers, each of which is defined by a set of $n_\ell$ binary variables, e.g. $\bs_{\ell}\in\{\pm 1\}^{n_\ell}$. Hence the number of states in layer $\ell$ is $2^{n_\ell}$, $\ell=1,2,\ldots$. 
Subsequent layers are connected by interaction parameters, and the first layer is connected to the "visible" layer $\bs_0\in\mathbb{R}^{n_0}$, which is composed of a high-dimensional vector ($n_0\gg 1$). 

This architecture corresponds to a joint probability distribution, for the visible ($\bs_0$) and the hidden ($\bs_\ell$) units, that reflects the Markov property of statistical dependencies between layers:
\begin{equation}
\label{joingen}
p(\bs_\ell,\bs_{\ell-1},\ldots,\bs_1,\bs_0)= p(\bs_0|\bs_1)\cdots p(\bs_{\ell-1}|\bs_\ell)p(\bs_\ell).
\end{equation}
In Eq.~(\ref{joingen}) 
\begin{equation}
\label{marggen}
p(\bs_\ell)=\sum_{\bs_{\ell+1},\ldots}p(\bs_\ell|\bs_{\ell+1})p(\bs_{\ell+1}|\ldots)\cdots
\end{equation}
is the marginal distribution of layer $\ell$, with respect to deeper layers $\ell+1,\ell+2,\ldots$. 

The interaction parameters between each layer are initialised to random values, and they are adjusted during the training phase in such a way as to maximise the likelihood of a dataset $\hat s_0=(\bs_0^{(1)},\ldots,\bs_0^{(N)})$ of a sample of $N$ data points. 

In an efficiently trained learning machine the statistical dependence needs to extend across different layers, in such a way that the state $\bs_\ell$ of a deep layer %when a particular input $\vec x$ is presented 
should significantly depend on $\bs_0$ (see Fig.~\ref{fig_OLMs} a). In addition, the learning machine should work also as a generative model. This means that 
the vector $\bs_0$ which can be generated propagating a state $\bs_\ell$ randomly drawn from $p(\bs_\ell)$ to the visible layer, should be statistically indistinguishable from the data points with which the machine was trained. In other words, the distribution $p(\bs_0)$ obtained from Eq.~(\ref{joingen}) marginalising over all hidden states should approximate the distribution of the data. 

Typical cases of interest are those where the data has a non-trivial structure. This is the case when the variation of $\bs_0$ across the samples spans a low dimensional manifold of intrinsic dimension $d$ which is much smaller than the number $n_0$ of components of $\bs_0$. Ansuini {\em et al.}~~\cite{ansuini2019intrinsic}, for example, estimate that the MNIST dataset ($n_0=784$) spans a space with intrinsic dimension $d\simeq 13$. In loose terms, learning entails extracting a compressed representation $\bs_\ell$ of the structure of statistical dependencies of the data. 

\subsection{Compression levels and the clamped distribution}

A natural measure for the level of compression is the entropy of the internal representation of the data. 
A proxy for this quantity is obtained considering the {\em distribution of clamped states} $q(\bs_\ell)$. This is obtained presenting each data point $\bs_0^{(i)}$ on the visible layer and finding the most likely corresponding {\em clamped} state 
\begin{equation}
\label{clamped}
\bs_\ell^*(\bs_0^{(i)})={\rm arg}\max_{\bs_\ell}p(\bs_\ell|\bs_0^{(i)})
\end{equation}
of layer $\ell$. $q(\bs_\ell)$ is the fraction of data points $\bs_0^{(i)}$ in the sample for which the clamped state coincides with $\bs_\ell$. The entropy 
\begin{equation}
\label{ }
\hat H[\bs_\ell]=-\sum_{\bs_\ell} q(\bs_\ell)\log q(\bs_\ell)
\end{equation}
of the distribution of clamped states provides a measure of the compression level achieved in layer $\ell$. 
When $p(\bs_\ell|\bs_0^{(i)})$ is a sharply peaked distribution, as is the case for well trained learning machines~\cite{SMJ,songthesis}, $q(\bs_\ell)$ is representative of the distribution obtained sampling the distribution $p(\bs_\ell|\bs_0)$ of internal states, when $\bs_0$ is a sampled datapoint. In addition, a well trained machine is expected to reproduce the (unknown) distribution of the data by the model $p(\bs_0)$ that is obtained from Eq.~(\ref{joingen}) by marginalisation over the hidden layers. In these cases, we expect that $q(\bs_\ell)\approx p(\bs_\ell)$ approximates well the marginal distribution of $\bs_\ell$, and that the entropy $\hat H[\bs_\ell]$  provides an estimate of the mutual information $I(\bs_\ell,\bs_0)=H[\bs_\ell]-H[\bs_\ell|\bs_0]$ between the data and the internal representation, because $\hat H[\bs_\ell]\approx H[\bs_\ell]$ and $H[\bs_\ell|\bs_0]\approx 0$.

 A plot of $\hat H[\bs_\ell]$ as a function of layer depth $\ell$ is shown in Fig.~\ref{fig_OLMs} b) for a DBN at different stages of training on the MNIST, for a DBN trained on the Fashion MNIST datasets and on randomised MNIST data\footnote{In randomised datasets pixels are reshuffled within each image.}. As Fig.~\ref{fig_OLMs} b) shows, the different layers of a well trained machine cover uniformly the range of entropies $H[\bs]\le \log N$, whereas  untrained DBNs generate representations whose entropy is either close to the entropy of the data ($\log N$) or to zero. A DBN trained on a randomised dataset exhibits a similar behaviour, which is consistent with the fact that the only structure left in the dataset is the variation in grey level across the sample.

\section{A random energy theory of deep learning}
\label{sec:remfit}

In order to discuss the statistical mechanics properties of the DBN, we consider it as a system in thermal equilibrium with a heat bath at inverse temperature $\beta=1$. Hence the Hamiltonian of the system is obtained by taking the logarithm of the joint distribution in Eq.~(\ref{joingen})
\begin{eqnarray}
\mathcal{H}(\bs_0,\ldots,\bs_\ell) & = & -\log p(\bs_0,\ldots,\bs_\ell) \\
 & = & u_{\bs_0|\bs_1}+\ldots +u_{\bs_{\ell-1}|\bs_\ell}+E_{\bs_{\ell}}-E_0.
 \label{REMass}
\end{eqnarray}
Apart for a constant $E_0$, which plays no major role in what follows, the Hamiltonian is a 
sum of a term $E_{\bs_\ell}=E_0-\log p(\bs_\ell)$ for layer $\ell$ and of interaction terms 
$u_{\bs_{k-1}|\bs_k}=-\log p(\bs_{k-1}|\bs_k)$ between layers, for $k=1,\ldots,\ell$. 

For a particular task, each of these terms takes a value which depends on the values of the weights with which the network is initialised and -- for a trained network -- on the dataset $\hat\bs_0$ with which the DBN is trained. 
In order to shed light on the generic nature of the energy landscape, we will compare the behaviour of 
learning machines with that of models where the terms in Eq.~(\ref{REMass}) are taken at random from a distribution, as in random energy models~\cite{derrida1981random}. As for models of spin glasses, the random energy assumption is questionable if taken too seriously. Yet, most of the results only depend on the degeneracy of energy levels (i.e. the number of states in a given energy range) so we believe that the conclusions that we shall derive provide a rather robust qualitative description of the statistical mechanics of probabilistic (deep) learning machines.

In what follows, we shall first discuss the properties of the internal layer $p(\bs_\ell)$ of a learning machine and then discuss the interaction with its environment, which is formed of the data and of the shallower layers.

\subsection{Random energy description of a single layer}

Let us first focus on the statistics of a generic hidden layer $\bs_\ell$. We shall drop the index $\ell$ in this subsection. 
In analogy with the REM, we assume that the energies $E_{\bs}$ are drawn  from the distribution
\begin{equation}
\label{Pu}
P\{E_{\bs}\le -\Delta z\}=e^{-z^\gamma}\,,\qquad z\ge 0
\end{equation}
independently for each state $\bs$. The statistical properties of the learned model are parametrised by $\Delta>0$ and $\gamma$, and they have been discussed in Ref.~\cite{derrida1981random,mezard2009information} for $\gamma=2$ and in Ref.~\cite{marsili2019peculiar} for $\gamma>0$, to which we refer for technical details. The derivation of the main results is discussed in the Appendix. We shall confine our discussion to the case $\gamma\ge 1$, since it can be argued~\cite{marsili2019peculiar} that the properties of the REM with $\gamma<1$ are not consistent with those of a system that learns. 

\begin{figure}
  \centering
\includegraphics[width=0.7\textwidth,angle=0]{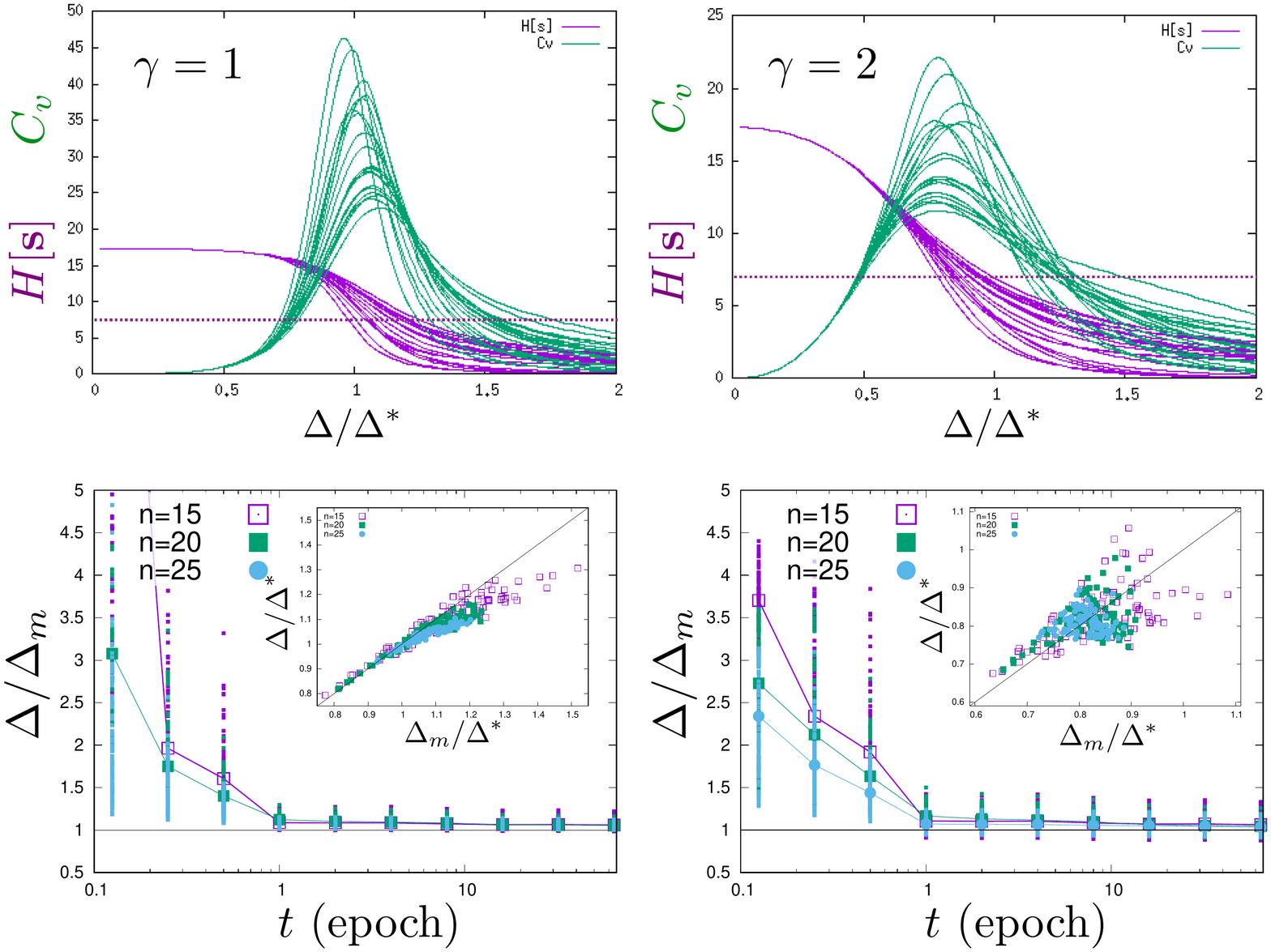}
  \caption{\label{fig_finitesize} Top row: Entropy $H[\bs]$ (purple lines) and specific heat $C$ (green lines) for 20 realisations of the REM with $n=25$ for $\gamma=1$ (left) and $\gamma=2$ (right). The dotted lines denote the values of $\hat H[\bs]$ for layer $\ell=6$ of the DBN of Fig.~\ref{fig_OLMs} trained on the MNIST dataset ($n=n_\ell=25$ nodes). Bottom row: Value of $\Delta$ obtained matching the observed value of $\hat H[\bs_\ell]$ for layers $\ell=6,7$ and $8$ of a DBN trained on MNIST ($n_\ell=25, 20$ and $15$, respectively).}
\end{figure}

The generic behaviour of the system in the thermodynamic limit, i.e. for $n\gg 1$, is characterised by a phase transition as $\Delta\to\Delta^*$, where 
\begin{equation}
\label{deltastar}
\Delta^*=\gamma\left(n\log 2\right)^{1-1/\gamma}.
\end{equation}
For $\Delta<\Delta^*$ the system exhibits a "high temperature" phase  where the entropy %\red{(details?)}
\begin{equation}
\label{Hsann}
H[\bs]\simeq\left[1-\left(\frac{\Delta}{\Delta^*}\right)^{\frac{\gamma}{\gamma-1}}\right]n\log 2
\end{equation}
is proportional to the number $n=n_\ell$ of nodes (see~\cite{marsili2019peculiar} for details). The entropy $H[\bs]$ instead remains finite as $n\to\infty$ for $\Delta>\Delta^*$. The transition is continuous but it becomes sharper and sharper as $\gamma\to 1^+$. 
For finite systems, $H[\bs]$ exhibits a variation as a function of $\Delta$ that become sharper and sharper in the neighbourhood of $\Delta^*$ as $n$ increases. The specific heat $C$, which is given by the variance of $E_{\bs}$, reaches a maximum at a value $\Delta_m$ which approaches $\Delta^*$ as $n$ grows large. The behaviour of $H[\bs]$ and $C$ as a function of $\Delta$ for $20$ realisations of the REM with $\gamma=1$ and $\gamma=2$ are shown in Fig.~\ref{fig_finitesize} (top) for $n=25$. 

\begin{figure}
  \centering
\includegraphics[width=0.7\textwidth,angle=0]{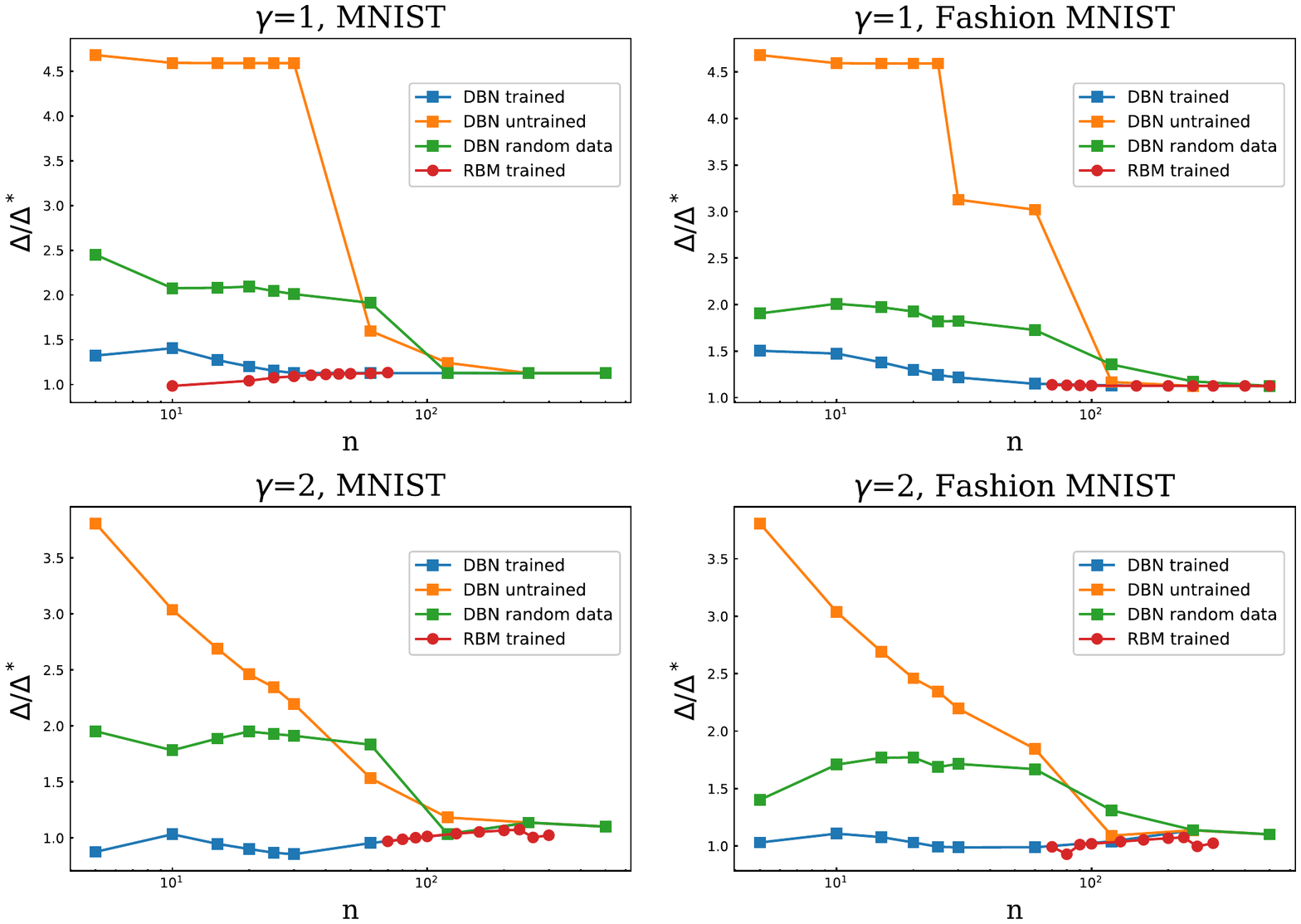}
  \caption{\label{fig_annealed} Estimate of the parameter $\Delta$ for $\gamma=1$ (top) and $\gamma=2$ (bottom) for a DBN with 10 layers with $n=500, 250, 120, 60, 30, 25, 20, 15, 10$ and $5$ units. Each panel shows the results for an untrained DBN, for DBN trained on the MNIST (left) and on Fashion MNIST (right) datasets, and for a DBN trained on shuffled datasets. Data for RBMs with $n\in [10,70]$ units trained on MNIST are also shown (see Appendix for more details).}
\end{figure}

In order to relate the behaviour of learning machines with REMs, we seek for the value of $\Delta$ for which the entropy $H[\bs]$ of a REM with $n$ variables equals the entropy $\hat H[\bs_\ell]$ of the layer of the DBN with the same number $n=n_\ell$ of variables (see Fig.~\ref{fig_finitesize} top).
Fig.~\ref{fig_finitesize} (bottom) shows the values of $\Delta$ obtained in this way for the layer with $n_\ell=25$ units of the DBN used in Fig.~\ref{fig_OLMs} (see caption) during training on the MNIST database. As this figure shows, the parameter $\Delta$ which best describes the distribution of energy levels of the internal representation approaches rapidly a value very close to $\Delta_m$, irrespective of the value of $\gamma$. For larger values of $n$, we resort to the annealed approximation of Ref.~\cite{marsili2019peculiar} to estimate $H[\bs]$ for the model, and compute the value of $\Delta$, matching the observed entropy $\hat H[\bs_\ell]$ for each layer, as explained above. The results are shown in Fig.~\ref{fig_annealed} for both DBMs and RBMs, for different dataset (see caption for details). We see that internal representations of learning machines trained on highly structured datasets are best described by REMs close to the critical point, i.e. $\Delta\simeq \Delta^*$. Untrained learning machines and machines trained on structureless datasets are instead described by off critical REMs with $\Delta$ that differs substantially from $\Delta^*$. For the shallower layers ($n\ge 10^2$) the variation of the REM's entropy is concentrated in a narrow interval around $\Delta^*$ while the entropy of the layer approaches the upper bound $\hat H[\bs_\ell]\le \log N$. These two effects conspire to give $\Delta\simeq\Delta^*$ even for untrained networks.

%For $\gamma=1$ the annealed approximation of Ref.~\cite{marsili2019peculiar} provides a rather simple equation for the entropy of the REM, which reads
%\begin{eqnarray}
%\label{Hsng1}
%H[\bs] & = & \frac{\Delta n\log 2}{1-2^{(\Delta-1)n}}+\frac{\Delta}{\Delta-1}+\log\frac{1-2^{(1-\Delta)n}}{\Delta-1} \\
% & \simeq & \frac{\log 2}{2}n+\log(n\log 2)+\frac{(n\log 2)^2}{12}(\Delta-\Delta^*)+\ldots\qquad (\Delta\approx\Delta^*=1)
% \nonumber
%\end{eqnarray}
%This equation can be used to obtain a rough estimate of the number of nodes needed to describe a dataset of $N$ points. Indeed, setting $\Delta=1$ and $H[\bs]=\log N$, Eq.~(\ref{Hsng1}) can be solved for $n$. For the MNIST dataset ($N=6\cdot 10^4$) this procedure predicts $n\simeq 23.67$. This is not very far from the number of hidden units ($n_\ell=25$) of the layer that best reproduces the statistics of the data, as discussed in Ref.~\cite{SMJ}, or the number of hidden units of the RBM ($n=\ldots$) or of the multi layer perceptron ($n=24$) which best describes the MNIST dataset~\cite{thesis_juyong}. Note, in particular, that the recipe suggested by Eq.~(\ref{Hsng1}) for choosing the number of hidden nodes differs substantially from the one proposed by Hinton~\cite{hinton2012practical}: while Eq.~(\ref{Hsng1}) suggests that $n$ should grow as $2\log N$ for $N$ large, Ref.~\cite{hintonguide} suggests a linear dependence $n\propto N$, with a coefficient that depends on the dimensionality of the data.

%\subsection{The criticality condition and the peak in the specific heat}

The critical point in the REM is located in the vicinity of the value $\Delta_m$ where the specific heat $C$ attains its maximum (see Fig.~\ref{fig_finitesize}). Tka\v{c}ik {\em et al.}~\cite{retina} suggest that this criterium can be advocated to detect criticality in general, given the estimate $p(\bs)$ of the probability distribution over the states of a system. This construction (see also~\cite{BialekUSSC}) is based on identifying the energy levels $E_\bs = -\log p(\bs)$ from the probability distribution, and analysing the statistical mechanical properties of a system whose partition function $Z(\beta)=\sum_{\bs} e^{-\beta E_\bs}$ is derived introducing a fictitious temperature $\beta$. In analogy with systems undergoing second order phase transitions, Tka\v{c}ik {\em et al.}~\cite{retina} identify the occurrence of a peak in the specific heat at $\beta=1$ as a signature of criticality of the system described by $p(\bs)$  (the activity of assembles of neurons in~\cite{retina}). We observe that this construction explores the large deviation properties of the system, where the expected value of $E_\bs$, which is the entropy $H[\bs]$, attains atypically small ($\beta>1$) or large ($\beta<1$) values\footnote{This is not true when the construction is based on a distribution $p(\bs)$ of sampled states, as in Refs.~\cite{retina,BialekUSSC}. This is because $p(\bs)$ only contains information on the typical behaviour at $\beta=1$ and it cannot describe large deviations.}. Hence this construction singles out the compression level $H[\bs]$ as the order parameter of the transition. This is a natural choice in learning systems, that are built in order to extract the most compressed representation of high dimensional datasets. Cubero {\em et al.}~\cite{RyanMDL} show that this same construction implies that efficient codes in Minimum Description Length (MDL)~\cite{MDL} are critical, thereby clarifying the nature of the phase transition. As in the REM, the critical state in MDL codes separates a "high temperature" phase of noisy representations from a "low temperature" one, which is dominated by just one state~\cite{RyanMDL}. This is reminiscent of the 
{\em mode collapse} phenomenon observed in generative adversarial networks~\cite{goodfellow2014generative}, which refers to the situation where the learned model ``specialises'' to 
generate only a limited variety of the inputs with which it has been trained. 
In this perspective, criticality of the internal representation arises as a consequence of learning machines striking the optimal trade-off between the accuracy with which it reproduces data points and its ability to generate the full variability of the data.

\begin{figure}
  \centering
\includegraphics[width=0.6\textwidth,angle=0]{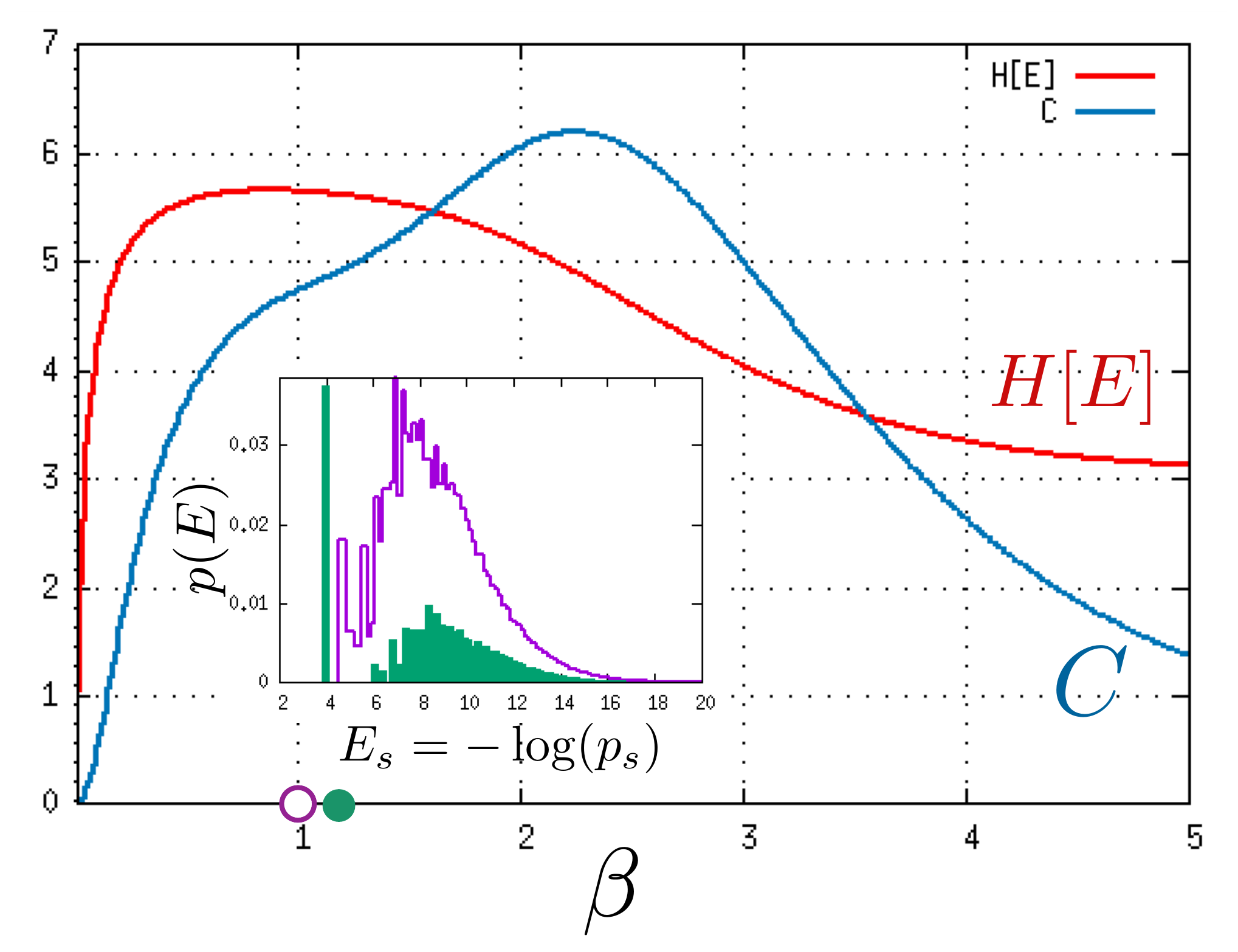}
  \caption{\label{fign22} Specific heat $C$ and relevance $H[E]$ as a function of the (fictitious) inverse temperature of the construction of Refs.~\cite{retina,BialekUSSC}, for a RBM with $n=22$ nodes trained on the MNIST database. The original distribution of energy levels $p(E)$ (i.e. for $\beta=1$) is compared to the one for $\beta=1.2$ in the inset. This shows that already close to the phase transition $p(\beta)$ develops a bimodal character, which is enhanced as $\beta$ increases. Close to the maximum of $C$, the gap between the ground state and the first excited state is of order $\Delta E\approx 10$.}
\end{figure}

We observe, however, that the location of the phase transition does not always coincide with a peak in the specific heat. In order to show this, Fig.~\ref{fign22} reports the values of the specific heat $C$ as a function of $\beta$ obtained with this construction, using the energy levels obtained from the internal representation $p(\bs)$ of a RBM with $n=22$ hidden units trained on the MNIST dataset. As Fig.~\ref{fign22} shows, $C$ exhibits a maximum at $\beta\approx 2.23$ rather than at $\beta=1$. This is a consequence of the fact that for $\beta>1$ the distribution $p(E)$ of energies develops a gap between the ground state and the bulk of excited states (see inset of Fig.~\ref{fign22}). This is fully consistent with the {\em mode collapse} phase transition discussed above. Fig.~\ref{fign22} also shows that the (differential) entropy $H[E]$ of the pdf of energies shows instead a maximum close to $\beta=1$. This suggests that $H[E]$ -- which has been called the {\em relevance} in~\cite{statcrit,odilon} -- is a better predictor of the location of the phase transition in learning systems. Both $C$ and $H[E]$ are measures of the width of the distribution of energy levels, but $C$ is maximal for bimodal energy distributions, whereas $H[E]$ is maximal when $p(E)$ covers densely a wide energy spectrum.

\subsection{The relation between layers}

The relation between internal states $\bs_\ell$ and inputs $\bs_0$ can be studied by considering the most likely {\em clamped} states $\bs^*_\ell,\bs^*_{\ell-1},\ldots\bs^*_1$ that the network associates to an input $\bs_0$. These are the solution of the maximisation problem 
\begin{equation}
\label{optmax}
\bs^*_\ell (\bs_0)={\rm arg}\max_{\bs_\ell}\left\{\log p(\bs_\ell)+\max_{\bs_{\ell-1}}v_{\bs_{\ell-1}|\bs_\ell}(\bs_0)\right\}
\end{equation}
%
%\begin{eqnarray}
%\bs^*_\ell (\vec x)& = & {\rm arg}\max_{\bs_\ell}\left\{\log p(\bs_\ell)+\max_{\bs_{\ell-1}}\left[\log p(\bs_{\ell-1}|\bs_\ell)+\max_{\bs_{\ell-2}}\left(\ldots+
%\max_{\bs_1}\log [p(\vec x|\bs_1)p(\bs_1|\bs_2)]\right)\right]\right\} \nonumber\\
% & = & {\rm arg}\max_{\bs_\ell}\left\{u_{\bs_\ell}+\max_{\bs_{\ell-1}}v_{\bs_{\ell-1}|\bs_\ell}(\vec x)\right\}\,,
% \label{optmax}
%\end{eqnarray}
where 
%we used Eq.~(\ref{ps}) for layer $\ell$ and 
we defined 
\begin{equation}
\label{leftsyst}
 v_{\bs_{\ell-1}|\bs_\ell}(\bs_0) = \max_{\bs_1\ldots,\bs_{\ell-2}}\log p(\bs_{\ell-1},\ldots,\bs_{1},\bs_0|\bs_\ell).
% \log p(\bt|\bs)+\max_{\bu}\left(\ldots+\max_{\bz}\log p(\bz|\vec x)\right)
\end{equation}
The term $v_{\bs_{\ell-1}|\bs_\ell}(\bs_0)$ characterises the ``environment'' with which the $\ell^{\rm th}$ internal layer interacts with, and it satisfies a recursion relation
\begin{eqnarray}
v_{\bs_{\ell-1}|\bs_\ell}(\bs_0)  & = & \log p(\bs_{\ell-1}|\bs_\ell)+\max_{\bs_{\ell-2}}\left[\max_{\bs_{\ell-3},\ldots,\bs_1}
\log p(\bs_{\ell-2},\bs_{\ell-3},\ldots,\bs_0|\bs_{\ell-1},\bs_{\ell})\right]\nonumber \\
 & = &   \log p(\bs_{\ell-1}|\bs_\ell)+\max_{\bs_{\ell-2}}\left[ \log p(\bs_{\ell-2}|\bs_{\ell-1})+v_{\bs_{\ell-2}|\bs_{\ell-1}}(\bs_0) \right]
 \label{prop}
 \end{eqnarray}
where we invoked the Markov property, i.e. the fact that $p(\bs_{\ell-2},\ldots,\bs_0|\bs_{\ell-1},\bs_{\ell})$ does not depend on $\bs_\ell$. Eq.~(\ref{prop}) describes how the statistical dependencies propagate across layers and Eq.~(\ref{optmax}) how it determines the distribution of clamped states
\begin{equation}
\label{pclamp}
q(\bs_\ell)=P\{\bs^*_\ell=\bs_\ell\}.
\end{equation}
Note that, if in some intermediate layer the optimisation on $\bs_{\ell-2}$ in Eq.~(\ref{prop}) is dominated by the first term $\log p(\bs_{\ell-2}|\bs_{\ell-1})$, then the dependence on $\bs_0$ of $v_{\bs_{\ell-1}|\bs_\ell}(\bs_0)$ is lost. This means that the solution of Eq.~(\ref{optmax}) will also be independent of $\bs_0$. In this case, the clamped distribution concentrates on just one state, which is what we observe in deep layers of untrained DBNs. On the other hand, if the optimisation in Eq.~(\ref{optmax})  is dominated by the term $v_{\bs_{\ell-1}|\bs_\ell}(\bs_0)$ each different input will likely result in a different internal layer $\bs_\ell^*$ and $\hat H[\bs_\ell]\simeq \log N$, as we observe in shallow layers of DBNs. These two extreme situations well characterise the behaviour of $H[\bs^*_\ell]$ observed in Fig.~\ref{fig_OLMs} for untrained learning machines.

Considering the input data $\bs_0$ as a random vector drawn from some (unknown) distribution implies that for all $\bs_\ell$ and $\bs_{\ell-1}$, $v_{\bs_{\ell-1}|\bs_\ell}$ is a random variable that we assume is drawn independently from a distribution
%\footnote{For simplicity we restrict to the case where $v_{\bt|\bs}\ge 0$. Strictly speaking this is inconsistent with the definition Eq.~(\ref{leftsyst}). Yet all the derivation generalise to the case where $v_{\bt|\bs}\to v_{\bt|\bs}+v_0$ is shifted by an arbitrary constant.}
\begin{equation}
\label{stretchedv}
P\{v_{\bs_{\ell-1}|\bs_{\ell}}>\Theta_{\ell}x-\bar v_\ell\}=e^{-x^{\theta_{\ell}}},\qquad (x\ge 0)\,,
\end{equation}
where $\theta_\ell\ge 1$, $\Theta_{\ell}$ sets the scale of the fluctuations of $v_{\bs_{\ell-1}|\bs_{\ell}}$ and $\bar v_\ell>0$ is a constant that plays no role in what follows\footnote{$\bar v_\ell$ should be large enough so that $v_{\bs_{\ell-1}|\bs_{\ell}}\le 0$ for all $\bs_{\ell-1}\in\{\pm 1\}^{n_{\ell-1}}$ and $\bs_{\ell}\in\{\pm 1\}^{n_{\ell}}$.}. Eq.~(\ref{stretchedv}) approximates the environment with which the $\ell^{\rm th}$ layer interacts with a REM.

The distribution of clamped states can be computed as in Ref.~\cite{marsili2019peculiar} using results from extreme value theory~\cite{Galambos}. Specifically, the intermediate maximisation of $v_{\bs_{\ell-1}|\bs_\ell}$ over $\bs_{\ell-1}$ (see e.g. Eq.~\ref{optmax}) can be carried out explicitly, with the result
\[
\max_{\bs_{\ell-1}}v_{\bs_{\ell-1}|\bs_\ell}\simeq a_\ell+\frac{\Theta_{\ell}}{\Theta^*_{\ell}}\eta_{\bs_{\ell}}
\]
where $a_{\ell}=\Theta_{\ell}(n_{\ell-1}\log 2)^{1/\theta_\ell}-\bar v_\ell$ is a constant, $\eta_{\bs_{\ell}}$ is a random variable that follows a Gumbel distribution 
\begin{equation}
\label{Gumbel}
P\{\eta_\ell<x\}=e^{-e^{-x}}.
\end{equation}
and
\begin{equation}
\label{eqbm}
\Theta^*_{\ell}=\theta_\ell (n_{\ell-1}\log 2)^{1-1/\theta_\ell}
\end{equation}
coincides with the critical value of $\Delta^*$ for $n=n_{\ell-1}$ and $\theta_\ell=\gamma_\ell$ (see Eq.~\ref{deltastar}). Since the maximum is taken over $2^{n_{\ell-1}}$ random variables, the characterisation provided by extreme value theory is accurate even for moderate values of $n_{\ell-1}$. This result can be used to compute the distribution of clamped states $\bs^*_\ell$, as shown in Ref.~\cite{marsili2019peculiar}, i.e.
\begin{eqnarray}
q(\bs_\ell) & = & P\left\{\log p({\bs_\ell})+\max_{\bs_{\ell-1}}v_{\bs_{\ell-1}|\bs_\ell}\ge \log p({\bs_\ell '})+\max_{\bs_{\ell-1}}v_{\bs_{\ell-1}|\bs_\ell'},~\forall\bs_\ell'\right\} \nonumber\\
& = & P\left\{\eta_{\bs_\ell'}\le \eta_{\bs_\ell}+\beta\log\frac{p({\bs_\ell})}{p({\bs_\ell'})},~\forall\bs_\ell'\right\} \label{GumbelEVT}\\
 & = & \frac{1}{Z(\beta)}p({\bs_\ell})^\beta,\qquad Z(\beta)=\sum_{\bs_\ell}p({\bs_\ell})^\beta\,,
\label{maxGB}
\end{eqnarray}
where we used Eq.~(\ref{Gumbel}) to compute Eq.~(\ref{GumbelEVT}) (see Appendix for more details). The exponent $\beta$ is given by
%This is a Gibbs-Boltzmann distribution with ``inverse temperature''
\begin{equation}
\beta=\frac{\Theta^*_{\ell}}{\Theta_{\ell}}=\frac{\theta_\ell}{\Theta_{\ell}} (n_{\ell-1}\log 2)^{1-1/\theta_\ell}
\end{equation}
that depends on the size $n_{\ell-1}$ of layer $\ell-1$ and on the parameter $\Theta_{\ell-1}$. Eq.~(\ref{maxGB}) shows that the distribution of clamped states $q(\bs_\ell)$ coincides with the distribution $p(\bs_\ell)$ only when $\beta= 1$, i.e. when the parameter $\Theta_{\ell}$ is tuned to the parameter $\Theta^*_{\ell}$ for which a REM with energies drawn from Eq.~(\ref{stretchedv}) is critical. 
In other words, statistical dependence on the data $\bs_0$ can propagate to deep layers of a neural network only if the environment with which layer $\ell$ interacts is tuned at the critical point, i.e. $\Theta_{\ell}\approx\Theta^*_{\ell}$, irrespective of the value of $\theta_\ell$. 
If $\Theta_\ell>\Theta^*_\ell$ the maximisation in Eq.~(\ref{optmax}) is dominated by the term $v_{\bs_{\ell-1}|\bs_\ell}$ which results in a noisy clamped distribution $q(\bs_\ell)$ which is broader than $p(\bs_\ell)$. If instead $\Theta_\ell\le \Theta^*_\ell$ the maximisation is dominated by the first term and $q(\bs_\ell)$ concentrates on the most probable values of $\bs_\ell$, i.e. those with higher $p(\bs_\ell)$. 

Let us discuss the implications of these findings on the performance of the learning machines as a generative model. 
Taking $q(\bs_\ell)$ as a projection of the data distribution in the $\ell^{\rm th}$ layer, the case $\beta<1$ corresponds to the situation where $p(\bs_\ell)$ is more sharply peaked than $q(\bs_\ell)$. Hence the DBN will not reproduce the full variability of the data with which it has been trained. Rather it will predominantly generate the most likely patterns. This is what the authors of Ref.~\cite{SMJ} observed when generating data points from deep layers of a DBN (see Fig.~\ref{fig_OLMs} b). 
Conversely, for $\beta<1$ the distribution $p(\bs_\ell)$ is flatter that $q(\bs_\ell)$. Noisy patters, which are generally unlikely in the data, are generated more frequently in this case. This behaviour is characteristic of shallow layers of DBN, as shown in Ref.~\cite{SMJ} (see also Fig.~\ref{fig_OLMs} b).

\section{Conclusion}
\label{sec:end}

Summarising, we study deep belief networks within a random energy ensemble approach. Each layer is described as a random energy model with a stretched exponential distribution of energies with parameter $\gamma$, as in Ref.~\cite{marsili2019peculiar}. Each layer features a phase transition to a ``frozen'' state for a particular value $\Delta^*$ of the interaction strength, which depends on the number of hidden units and on $\gamma$ (Eq.~\ref{eqbm}). We show that {\em i)} the internal representation of hidden layers of well trained learning machines (DBN and RBM) on structured datasets (MNIST and Fashion MNIST) are best described by REMs with a parameter that is close to the critical point $\Delta^*$, irrespective of $\gamma$ in a wide range of $n$. Untrained learning machines or machines trained on structureless data are instead best fitted by off-critical REMs. Furthermore, {\em ii)} in order to propagate the dependence on the data to the deep layers, each layer should interact with the data through a system (the environment) which should be tuned at the critical point of the corresponding REM. 

Notice that close to the critical point $\Delta^*$, the distribution of energies of a REM is approximately flat, irrespective of the value of $\gamma$ (see the Appendix). This prediction can be tested empirically, because, together with the relation between the clamped distribution and the distribution $p(\bs)$ of Eq.~(\ref{maxGB}), it implies that the distribution of clamped states should exhibit statistical criticality. Refs.~\cite{hennig,SMJ} provide ample evidence that support this claim.

\section{Acknowledgments}

Rongrong Xie acknowledges a fellowship from the China Scholarship Council (CSC) under Grant CSC No. 202006770018.
 
\appendix

\section{The generalised REM}

The thermodynamic properties of the REM are derived following the same arguments as in Ref.~\cite{derrida1981random,mezard2009information,marsili2019peculiar}. The behaviour in the random phase ($\Delta<\Delta^*$) can be discussed within the annealed approximation. The number of states at energy $E$ is asymptotically given by 
$e^{S(E)}$ where the thermodynamic entropy is
\begin{equation}
\label{ }
S(E)=n\log 2 -\left(- E/\Delta\right)^\gamma\,.
\end{equation}
The ground state energy is given by the value for which $S(E_{GS})=0$. This yields $E_{GS}=\Delta (n\log 2)^{1/\gamma}$. For small values of $\Delta$, the partition function
\begin{equation}
\label{ }
Z(\Delta)=\sum_{\bs} e^{-E_{\bs}} \simeq \int dE e^{S(E)-E}
\end{equation}
is dominated by the saddle point value $E(\Delta)$ for which $\frac{dS}{dE}=1$. This gives
\begin{equation}
\label{Eann}
E(\Delta)=-\Delta\left(\Delta/\gamma\right)^{1/(\gamma - 1)}.
\end{equation}
This expression is valid as long as $E(\Delta)> E_{GS}$. The critical point is given by the point where the thermodynamics becomes dominated by few states with energy close to the ground state, i.e. when $E(\Delta^*)= E_{GS}$. This condition yields Eq.~(\ref{deltastar}). In order to obtain Eq.~(\ref{Hsann}) we observe that Eq.~(\ref{Eann}) can be written as $E(\Delta)=E_{GS}(\Delta/\Delta^*)^{1/(\gamma-1)}$. This readily gives the expression Eq.~(\ref{Hsann}) for the entropy $H[\bs]=S\left(E(\Delta)\right)$. At $\Delta^*$, the expansion of the entropy around $E_{GS}$ yields 
\[
S(E)\simeq E-E_{GS}+\frac{\gamma-1}{2}\frac{(E-E_{GS})^2}{E_{GS}}+O((E-E_{GS})^3)
\]
which is approximately linear on an energy range $\delta E\ll \sqrt{-E_{GS}}=\sqrt{\gamma n\log 2}$, for $\gamma>1$. Hence the distribution of energies $p(E)\simeq e^{S(E)-E}$ extends over a range of order $\delta E\sim \sqrt{n}$ and the specific heat $C\sim\delta E^2$ is proportional to $n$. For $\gamma=1$ instead the entropy is linear in $E$ for all values of $\Delta$ and at $\Delta^*$ the distribution of energies extends over a range proportional to $n$, so that $C\sim n^2$.

In order to derive Eq.~(\ref{maxGB}), we observe~\cite{marsili2019peculiar} that the probability in Eq.~(\ref{GumbelEVT}) can be evaluated explicitly using Gumbel distribution for $\eta_{\bs}$. Indeed
\begin{eqnarray}
P\left\{\eta_{\bs_\ell'}\le \eta_{\bs_\ell}+z_{\bs|\bs'},~\forall\bs'\right\} & = & 
\int_{-\infty}^\infty \! d\eta_{\bs}e^{-\eta_{\bs}-e^{-\eta_{\bs}}}\prod_{\bs'\neq \bs} P\{\eta_{\bs'}\le\eta_{\bs}+z_{\bs|\bs'}\}
\nonumber\\
 & = & \int_0^{\infty}dy e^{-y-y\sum_{\bs'\neq\bs} e^{-z_{\bs|\bs'}}}
\end{eqnarray}
where we used the fact that $P\{\eta_{\bs'}\le\eta_{\bs}+z_{\bs|\bs'}\}=e^{-e^{-\eta_{\bs} - z_{\bs|\bs'}}}$ and changed variables to $y=e^{-\eta_{\bs}}$.

\section{Data and network architectures}

This study employs the handwritten digits dataset MNIST, and the article images dataset Fashion-MNIST.  Both databases, consisting of $28 \times 28$ grayscale images with 10 different classes. Both are split into a training sets with $60000$ images and a test sets with $10000$ images.

We adopted the Persistent Contrastive Divergence algorithm to train DBN and RBM. The training process runs for 64 epochs with a mini-batch size of $100$ and a learning rate of $0.1$. We set the persistent chains to be $100$ steps long, and the initial weights to be $4\sqrt{\frac{6}{nVis + nHid}} \times N(nVis, nHid)$,  where $nVis$ and $nHid$ are the number of visible units and hidden units, respectively,  and $N(nVis, nHid)$ is a sample generated from the standard normal distribution and with a dimension of $nVis \times nHid$. The initial visible and hidden biases are assigned to be zero with a dimension of $1 \times nVis$  and $1 \times nHid$,  respectively.

In the $\gamma=1$ case, as shown in the upper subplots in Figure 3,  we use a different number of hidden units in RMB when training with these two datasets.  For MINIST, we set it to be $10, 20, 25, 30, 35, 40, 45, 50, 60$ and $70$, respectively, while for Fashion MNIST, we fix it to be $70, 80, 90, 100, 150, 200, 250, 300, 400$ and $500$, respectively. In the $\gamma=2$ case, as shown in the lower subplots in Figure 3,  the number of hidden units in RBM is the same when training with MINIST and Fashion MNIST. Both are set to be $70, 80, 90, 100, 130, 160, 200, 230, 260$ and $300$, respectively.  For the untrained DBN,  we use the initial parameters (weight, visible bias and hidden bias) instead of using the well-trained parameters to compute in order to  $\hat H[\bs]$ and the corresponding value of $\Delta$.  For the DBN random data, we purposely randomise the positions of pixels within each image in both MNIST and Fashion MNIST.

%%%%%%%%%%%%%%%%%%%%%%%%%%%%%%%%%%%%%%%%%%%%%%%%%%%%%%%%%%%%

\bibliographystyle{plain} 
\bibliography{neurips_2021.bib}

\end{document}